\documentclass[12pt]{article}

\usepackage{latexsym,amsmath,amssymb,theorem,epsfig}

\DeclareFontFamily{OT1}{rsfs10}{}
\DeclareFontShape{OT1}{rsfs10}{m}{n}{ <-> rsfs10 }{}
\DeclareMathAlphabet{\mathscript}{OT1}{rsfs10}{m}{n}

\numberwithin{equation}{section}

\newcommand{\ns}{\normalsize}

\def\a{\alpha}

\theoremstyle{plain}

{\theorembodyfont{\rmfamily} }

\begin{document}

%%%%%%%%%%%%%%%%%%%%%%%%%%%%%%%%%%%%%%%%%%%%%%%%%%%%%%%%%%%%%%%%%%%%%%

\begin{titlepage}

\vspace{-5cm}

\title{
  \hfill{\ns }  \\[1em]
   {\LARGE On Open Membranes, Cosmic Strings and Moduli Stabilization}
\\[1em] }
\author{
   Evgeny I. Buchbinder
     \\[0.5em]
   {\ns School of Natural Sciences, Institute for Advanced Study} \\[-0.4cm]
{\ns Einstein Drive, Princeton, NJ 08540}\\[0.3cm]}

\date{}

\maketitle

\begin{abstract}

We discuss how cosmic strings can be created in heterotic M-theory
compactifications with stable moduli. We conclude that the only
appropriate candidates seem to be fundamental open membranes with
a small length. In four dimensions they will appear as strings
with a small tension. We make an observation that, in the
presence of the vector bundle moduli, it might be possible to
stabilize a five-brane very close to the visible sector so that a
macroscopic open membrane connecting this five-brane and the
visible brane will have a sufficiently small length. We also
discuss how to embed such cosmic strings in heterotic models with
stable moduli and whether they can be created after inflation.

\end{abstract}

\thispagestyle{empty}

\end{titlepage}

%%%%%%%%%%%%%%%%%%%%%%%%%%%%%%%%%%%%%%%%%%%%%%%%%%%%%%%%%%%%%%%%%%%%%%%%%%%%%%%%%%%%%%%%%%%%%%%%%%%%%%%%

\section{Introduction}

%%%%%%%%%%%%%%%%%%%%%%%%%%%%%%%%%%%%%%%%%%%%%%%%%%%%%%%%%%%%%%%%%%%%%%%%%%%%%%%%%%%%%%%%%%%%%%%%%%%%%%%%%

String theory seems to have appropriate candidates for cosmic
strings. The possibility of viewing superstrings as cosmic
strings was first discussed by Witten in~\cite{Wittencosmic}. He
pointed out some serious problems in this direction. For example,
one of the common problems is that the tension of fundamental
superstrings is much bigger than the observational bound on the
cosmic string tension. Another very common problem is that cosmic
superstrings appear as boundaries of axion domain walls. This
will force the strings to collapse after the axion gets a
mass~\cite{Vilenkin}.

Cosmic strings in the context of string theory were recently
reconsidered in~\cite{Tye1, Tye2, Dvali, CMP}. It was argued that
the previous problems might be avoided thanks to recent
developments in string theory, including D-, NS- and M-branes,
compactifications with large warp factors and models with large
extra dimensions. In particular, small tension of cosmic
superstrings can be explained by a large warp factor. A
systematic analysis of fundamental and Dirichlet cosmic strings
was performed by Copeland, Myers and Polchinski in~\cite{CMP}.
Results obtained in~\cite{CMP} suggest that in type IIB
compactifications (at least in models with many
Klebanov-Strassler throats~\cite{Klebanov}) it is possible to
have brane/anti-brane inflation~\cite{Kachruinfl} followed by
production of metastable F- and D-strings and de Sitter (dS)
vacua~\cite{KKLT}. Furthermore, it was proposed in~\cite{Dvali1,
Dvali2, Redi} that cosmic D-strings can be identified with
solitons in supergravity with Fayet-Iliopoulos (FI) terms. A
holographic dual of D-strings in the throat was studied
in~\cite{Gubser}.

The most phenomenologically attractive four-dimensional vacua
arise in string/M-theory from compactifications of strongly
coupled heterotic string~\cite{HW1, HW2, Wittenstrong}. Recently,
it was demonstrated~\cite{He1, He2} that heterotic string theory
on smooth non-simply connected Calabi-Yau manifold with $Z_3
\times Z_3$ homotopy group can give rise to a four-dimensional
theory whose spectrum differs form the spectrum of the Standard
Model only by one extra Higgs multiplet. A progress has also been
made in heterotic models towards moduli
stabilization~\cite{Curio, BO, Gukovhet, Becker, Raise} and
inflation~\cite{Myinflation, Theirinflation}. In this paper, we
ask a question whether it is possible to find cosmic strings in
heterotic compactifications with stable moduli. In section 2, we
discuss several ways how cosmic strings could be created and come
to conclusion that the only candidates seem to be fundamental open
membranes with a small length. In four dimensions, an open
membrane appears as a string. Such membranes could arise if one
of five-branes is stabilized close to one of the orbifold fixed
planes. In section 3, we show that it might be possible to
stabilize a five-brane close to the visible brane due to its
non-perturbative interaction with the vector bundle moduli. In
section 4, we discuss stabilization of the remaining moduli. We
slightly generalize the earlier results~\cite{BO, Raise} and
point out that it should be possible to stabilize the interval in
the phenomenological range in the presence of many five-branes in
the bulk without any extra assumptions. Cosmic strings are
unstable in the presence of axion domain walls. In~\cite{CMP}, it
was shown that axion domain walls will not be formed if the axion
is charged under a $U(1)$ gauge group. In heterotic string theory
this happens if in four dimensions there is an anomalous $U(1)$
whose anomaly is canceled by the Green-Schwarz
mechanism~\cite{DSRW}. However, to keep this argument valid, it
is important that the axion does not receive any potential. Since
in heterotic M-theory the axion is the imaginary part of the
volume multiplet~\cite{Lukas4, Lukas5}, no superpotential for the
volume should be present. This makes it difficult to stabilize
the volume. Fortunately, the presence of the anomalous $U(1)$
implies the presence of volume-dependent FI-terms. We show that
together with the F-terms, the FI-terms naturally provide
stabilization of all moduli in a non-supersymmetric Anti de Sitter
(AdS) vacuum. In subsection 4.2, we speculate how cosmic strings
can be embedded into compactifications with dS vacua. We show
that it might be possible in non-supersymmetric
compactifications. In section 5, we discuss how the cosmic
strings, studied in this paper, can be created after inflation.

%%%%%%%%%%%%%%%%%%%%%%%%%%%%%%%%%%%%%%%%%%%%%%%%%%%%%%%%%%%%%%%%%%%%%%%%%%%%%%%%%%%%%%%%%%%%%%%%%%%%%%%%%%%

\section{Cosmic Strings in Heterotic M-theory}

%%%%%%%%%%%%%%%%%%%%%%%%%%%%%%%%%%%%%%%%%%%%%%%%%%%%%%%%%%%%%%%%%%%%%%%%%%%%%%%%%%%%%%%%%%%%%%%%%%%%%%%%%%

In this section, we will point out that it is harder to find
string-like objects in Calabi-Yau compactifications of heterotic
M-theory comparing to type II compactifications. Let us start with
discussing whether cosmic strings can arise after inflation
considered in~\cite{Myinflation}. The inflationary phase
in~\cite{Myinflation} is represented by a five-brane, wrapped on
an non-isolated genus zero or a higher genus holomorphic curve in
a Calabi-Yau threefold, approaching the visible
brane\footnote{Throughout the paper, we will refer to one of the
orbifold fixed planes as to the visible brane, or the visible
sector, and to the other one as to the hidden brane, or the hidden
sector.}. It was argued in~\cite{Myinflation} that as the
five-brane comes too close the transition vector bundle
moduli~\cite{BDO1, Rene} associated with this holomorphic curve
become tachyonic terminating inflation. In the post-inflationary
phase, the five-brane disappears (becomes massive) through the
small instanton transition~\cite{Wittensmall, Seiberg, OPP},
whereas the transition vector bundle moduli can be stabilized by
non-perturbative effects. Thus, the post-inflationary physics does
not contain any rolling moduli. All the remaining moduli are
stabilized both during and after inflation. The cosmological
constant, generically, changes after the small instanton
transition. This leads to the possibility of obtaining dS vacua
with a small cosmological constant in the post-inflationary phase.
A five-brane wrapped on a Riemann surface of genus one or higher
carries one or more $U(1)$ gauge fields on its
world-volume~\cite{Nonstandard}. For simplicity, consider a
five-brane wrapping a torus so that there is only one $U(1)$
factor. During the small instanton transition, this $U(1)$ gets
broken. Therefore, it seems conceivable that cosmic strings can be
created by the Kibble mechanism~\cite{Kibble}. However, it is not
the case. First problem is that the small instanton transition may
not be describable by field theory, whereas the Kibble argument is
based on the Lagrangian description. Second, even if there is a
field theory description of the small instanton transition (one
might expect it if a Calabi-Yau threefold is not simply
connected), one should recall that the theory we deal with is the
$S^1/Z_2$ orbifold. Therefore, when a five-brane coincides with
one of the orbifold fixed plains, it also coincides with its
image. In this case, one should expect that the gauge group which
gets restored is $SU(2)$. As a result, after the small instanton
transition the group which is broken is $SU(2)$ and no strings are
formed. This set-up has one more candidate for cosmic string in
the vector-bundle branch. When a five-brane coincides with an
orbifold fixed plane, a tensionless string arises~\cite{Hanany}.
Can it get a tension in the vector bundle branch, with its
tension being set by the vector-bundle moduli? The answer is again
negative. This tensionless string should be understood as the
analog of a monopole becoming light. This analogy suggests that
this string will get a tension in the five-brane branch. In the
vector bundle branch, it will not exist for the same reason why
monopoles do not exist in the Higgs branch. This discussion
demonstrates that no cosmic strings are formed after inflation
presented in~\cite{Myinflation}. Below, in the paper, we will
modify the set-up of~\cite{Myinflation} to allow cosmic strings
to be formed after inflation.

Let us now discuss whether it is possible to create cosmic strings
after a hypothetical (not yet constructed)
five-brane/anti-five-brane inflation. After
five-brane/anti-five-brane annihilation, a membrane bound state
can be formed~\cite{Yi}. This membrane will be parallel to the
orbifold fixed planes. If our Calabi-Yau threefold is non-simply
connected, one can wrap such a membrane on a one-cycle with
non-trivial $\pi_1$ and form a string. However, this string will
be stable in M-theory but not in heterotic M-theory. A membrane
parallel to the orbifold fixed planes is not
supersymmetric~\cite{Lima1, Lima2}. Therefore, one should expect
that it will quickly annihilate with its image.

Our discussion shows that it is hard to find cosmic strings in
heterotic M-theory. However, we have not yet considered the most
natural candidate, a fundamental open membrane. From the four-dimensional
viewpoint such a membrane looks like a string. The tension of such
a string behaves as
\begin{equation}
\mu \sim M_{11}^3y,
\label{1.1}
\end{equation}
where $M_{11}$ is the eleven-dimensional Planck scale and $y$ is
the length of the membrane in the eleventh dimension. It is easy
to estimate that in order to fulfill the observational bound $G\mu
< 10^{-7}-10^{-6}$ (see, for example,~\cite{Polchinski} and
references therein), the ratio of $y$ to the size of the interval
$\pi \rho$ should be of order
\begin{equation}
\frac{y}{\pi \rho} \sim 10^{-4}-10^{-3}.
\label{1.2}
\end{equation}
To obtain~\eqref{1.2}, we set the volume scale $v_{CY}$ and the
interval scale $\pi \rho$ to be in the phenomenological range,
$v_{CY}^{1/6} \sim (10^{16}{\ } GeV)^{-1}$ and $\pi \rho \sim
(10^{15}{\ } GeV)^{-1}$ and used the fact that the
eleven-dimensional and four-dimensional Planck scales, $M_{11}$
and $M_{Pl}$, are related as follows
\begin{equation}
M^{2}_{Pl}=M^{9}_{11}v_{CY}\pi \rho.
\label{1.3}
\end{equation}
Production of strings with such a tension implies some constraints
on the five-brane stabilization. One of the five-branes in the
bulk has to be stabilized very close to either one of the orbifold
fixed planes or another five-brane. In the next section, we will
argue that it is possible to stabilize a five-brane close to the
visible brane. The key role in this process is played by the
vector bundle moduli and their interaction. In later sections, we
will extend results of section 3 to include the remaining moduli
and the inflaton.

%%%%%%%%%%%%%%%%%%%%%%%%%%%%%%%%%%%%%%%%%%%%%%%%%%%%%%%%%%%%%%%%%%%%%%%%%%%%%%%%%%%%%%%%%%%%%%%%%%%%%%%%%%%

\section{Five-brane Stabilization and Open Membranes with a Small Length}

%%%%%%%%%%%%%%%%%%%%%%%%%%%%%%%%%%%%%%%%%%%%%%%%%%%%%%%%%%%%%%%%%%%%%%%%%%%%%%%%%%%%%%%%%%%%%%%%%%%%%%%%%%%%%

We consider a Calabi-Yau compactification of heterotic M-theory
with the following complex moduli
\begin{equation}
S, \quad T^{I}, \quad Y, \quad Z_{\a};\quad Y_0, \quad \phi_i.
\label{3.1}
\end{equation}
Here $S=V+i\sigma$, where $V$ is the volume modulus and $a$ is the
axion, $T^{I}$'s are the Kahler moduli. They are constructed as
follows~\cite{Lukas4, Lukas5}
\begin{equation}
T^{I}=Rb^{I}+ip^{I}, \quad I=1, \dots, h^{1,1},
\label{3.2}
\end{equation}
where $R$ is the interval modulus, $b^{I}$ are the $(1,1)$ moduli
of the Calabi-Yau threefold and $p^{I}$'s come from the
components of the M-theory three-form $C$ along the interval and
the Calabi-Yau manifold. The moduli $Y$ and $Y_0$ correspond to
five-branes wrapped on an isolated genus zero holomorphic curve
in the Calabi-Yau threefold. The real part of each of the two
five-brane multiplets is the position of the corresponding
five-brane in the interval. We denote them by $y$ and $y_0$
respectively. The imaginary part is related to the axions on the
five-brane worldvolume. See~\cite{Deren} for details. One
five-brane is needed to stabilize the interval in a
phenomenological range as in~\cite{BO}, \cite{Raise}. This
five-brane has to be located relatively close to the hidden
brane. The other five-brane will be the one that we will attempt
to stabilize in this section close to visible brane so that
eq.~\eqref{1.2} is satisfied. $Z_{\alpha}$ are the complex
structure moduli, whose actual number is irrelevant. We will
assume that the $(3,0,1)$ flux $G$ is turned on. Such a flux
produces the superpotential for $Z_{\alpha}$ of the
form~\cite{Gukov, Constantin}
\begin{equation}
W_{f}=\frac{M^2_{Pl}}{v_{CY}\pi\rho} \int dx^{11}\int_{CY} G\wedge \Omega.
\label{3.4}
\end{equation}
This superpotential is expected, generically, to stabilize all the
complex structure moduli. We will assume that it is the case.
We will also assume that the complex structure moduli are stabilized at slightly
higher scale that all the other moduli so that $W_f$ can be considered as constant
in the low-energy field theory.
At
last, $\phi_i$ are the moduli of the vector bundle on the visible
brane.
%Without loss of generality, we can assume that there is
%only one such a modulus $\phi$.
For simplicity, we will assume that the bundle on the hidden
brane either is trivial or does not have any moduli. The moduli
$V$ and $R$ are assumed to be dimensionless and normalized with
respect to the reference scales $v_{CY} \sim (10^{16}{\ }
GeV)^{-1}$ and $R \sim (10^{15}{\ }GeV)^{-1}$. The moduli $y$ and
$y_0$ are also dimensionless and less that $R$. To obtain the
four-dimensional coupling constants in a phenomenological range,
$V$ and $R$ have to be stabilized at (or be slowly rolling near)
a value of order one.

In this section, we will concentrate on the moduli $Y_0$ and
$\phi$ assuming that all the other moduli are stable. They will be
considered in the next section. The Kahler potential of the system
(in the four-dimensional Planck units) is given by
\begin{equation}
K(Y_0, \phi) =K_0+K_1(Y_0+{\bar Y}_{\bf 0})^2 +kK(\phi).
\label{3.5}
\end{equation}
Here $K_0$ and $K_1$ are functions of the remaining moduli
independent of $Y_0$ whose precise form is irrelevant for us.
$K(\phi)$ is the Kahler potential for the vector bundle moduli
$\phi_i$ and $k$ is a dimensionless parameter of order
$10^{-5}$~\cite{BO}. It is much less than one because the gauge
sector in Horava-Witten theory~\cite{HW1, HW2} arises at the
sub-leading order in the gravitational coupling constant
comparing to the gravity sector. The superpotential (in the
four-dimensional Planck units) is given by
\begin{equation}
W=W_0 + \alpha Pfaff({\cal D}_{-})e^{-\tau Y_0}. \label{3.7}
\end{equation}
Here the second term is the non-perturbative superpotential due
to a membrane instanton~\cite{Wittenpoten, BBS, Moore, Lima1,
Lima2, BDO2, BDO3}, with $\alpha$ being a dimensionless parameter,
and $W_0$ is the superpotential depending on the remaining moduli.
For our purposes it is a constant. The parameter $\tau$ is given
by
\begin{equation}
\tau =\frac{1}{2} (\pi \rho) v (\frac{\pi}{2 \kappa_{11}})^{1/3},
\label{3.7.1}
\end{equation}
where $v$ is the area of the holomorphic curve on which the
membrane is wrapped. Generically, it is of order $100$.
$Pfaff({\cal D}_{-})$ is the Pfaffian of the Dirac operator
depending on the vector bundle moduli. It is very difficult to
compute it explicitly because no gauge connections on Calabi-Yau
threefolds are known. However, in~\cite{BDO2, BDO3}, this factor
was computed in a number of examples for the case the Calabi-Yau
threefold elliptically fibered over the Hirzebruch surface using
algebraic geometry techniques. It was found to be a high degree
homogeneous polynomial in the vector bundle moduli. We will denote
it as ${\cal P}(\phi)$. If the five-brane is close to the visible
sector, all other $\phi_i$ and $Y_0$ contributions to the
non-perturbative superpotential are negligibly small. If we
ignore all other moduli (they will be discussed later), the
minimum of the potential energy for the $Y_0, \phi_i$ system is
given by
\begin{equation}
D_{\phi_i}W=0, \quad D_{Y_0}W=0, \label{3.9}
\end{equation}
where $DW$ is the Kahler covariant derivative, $DW=\partial W+
\frac{W}{M^2_{Pl}}\partial K$. From eqs.~\eqref{3.5} and
\eqref{3.7} we have
\begin{equation}
D_{\phi_i}W = \frac{\partial {\cal P}(\phi)}{\partial \phi_i}
e^{-\tau Y_0} +k \frac{\partial K(\phi)}{\partial \phi_i} W =0
\label{3.10}
\end{equation}
and
\begin{equation}
D_{Y_0}W = -\tau {\cal P}(\phi) e^{-\tau Y_0} +4K_1 y_0 W=0.
\label{3.11}
\end{equation}
It is easy to realize that since $\tau$ is much bigger than one
and $k$ is much less than one, in the second terms of
eqs.~\eqref{3.10} and~\eqref{3.11}, we can replace $W$ with $W_0$.
If the degree of the polynomial ${\cal P}(\phi)$ is sufficiently
high, the first terms is eqs.~\eqref{3.10} and~\eqref{3.11} are
practically the same. The second term in eq.~\eqref{3.10} is
proportional to $k$ whereas the second term in eq.~\eqref{3.11}
is proportional to $y_0$. These two equations are consistent only
if $y_0 \propto k$, that is $y_0$ is much less than one. Whether
or not the solution for $y_0$ is consistent with~\eqref{1.2}
depends on details of the compactification, in particular, on the
Kahler potential $K(\phi)$ which is unknown how to evaluate
explicitly on a Calabi-Yau threefold. Nevertheless, the form of
eqs.~\eqref{3.10} and~\eqref{3.11} suggests that in some models
it might be possible to find a solution for $y_0$ consistent
with~\eqref{1.2}.

To get a better understanding how a solution for $y_0$ might look
like, let us make a reasonable model for the Kahler potential
$K(\phi)$. Let us choose, for simplicity, $K(\phi)$ to be flat
\begin{equation}
K(\phi)=\sum_{i=1}|\phi_i|^2. \label{3.12}
\end{equation}
It corresponds to the Kahler potential for the size and
orientation moduli of centered Yang-Mills instantons  on ${\mathbb
R}^4$ if the number of moduli $\phi_i$ is even~\cite{Dorey,
Gray1, Gray2}. The Pfaffian will be assumed to be a generic
homogeneous polynomial of degree $d$. It is obvious that for a
generic Pfaffian it is possible to find a solution for the phases
of $\phi$'s and the imaginary part of $Y_0$. Let $r_i$ be the
absolute value of $\phi_i$. Consider eqs.~\eqref{3.10} for $r_i$.
Since we are interested in solutions with $\tau y_0 << 1$, the
exponential factor $e^{-\tau y_0}$ in eqs.~\eqref{3.10},
\eqref{3.11} can be set to unity. Then using eq.~\eqref{3.12},
eqs.~\eqref{3.10} for $r_i$ can be written as follows
\begin{equation}
{\cal F}_i(r_j)= \frac{k |W_0|}{\alpha d}, \label{nn1}
\end{equation}
where each ${\cal F}_i$ is a homogeneous function of $r_j$ of
degree $d-2$. The factor $d$ in the denominator in the left hand
side comes form differentiating the Pfaffian\footnote{Without
loss of generality, we can assume that the Pfaffian contains terms
$r_j^d$.}. From eqs~\eqref{nn1} it follows that every $r_i$ is of
the form
\begin{equation}
r_i \propto \left( \frac{k|W_0|}{\alpha d} \right)^{1/(d-2)},
\label{3.13}
\end{equation}
where the proportionality coefficient is a number depending on
details of the Pfaffian. Then from eq.~\eqref{3.11} it follows
that
\begin{equation}
y_0 \sim \frac{\alpha \tau}{4 K_1 |W_0|} \left( \frac{k
|W_0|}{\alpha d}\right )^{d/(d-2)}. \label{3.14}
\end{equation}
If $d >>1$, we have
\begin{equation}
y_0 \sim \frac{\tau k}{4 K_1 d}. \label{3.15}
\end{equation}
Taking, for example, $k\sim 10^{-5}, \tau \sim 100, d \sim 10,
4K_1\sim 1$, we find that
\begin{equation}
y_0 \sim 10^{-4}
\label{3.16}
\end{equation}
which is consistent with condition~\eqref{1.2}. Note that $y_0$
is inverse proportional to $d$. A large value of $d$, which is
consistent with results of~\cite{BDO2, BDO3}, might also be a
reason for a small value of $y_0$.

In this section, we demonstrated that it is possible to stabilize
a five-brane, wrapped on an isolated genus zero holomorphic
curve, close to the visible brane. From the four-dimensional
viewpoint, a macroscopic open membrane stretched between this
five-brane and the visible brane will look like a cosmic string
with a relatively small tension which might be consistent with
the observational bound.

%%%%%%%%%%%%%%%%%%%%%%%%%%%%%%%%%%%%%%%%%%%%%%%%%%%%%%%%%%%%%%%%%%%%%%%%%%%%%%%%%%%%%%%%%%%%%%%%%%%%%%%%%%%%%%%%%%%%

\section{Stabilization of the Remaining Moduli}

%%%%%%%%%%%%%%%%%%%%%%%%%%%%%%%%%%%%%%%%%%%%%%%%%%%%%%%%%%%%%%%%%%%%%%%%%%%%%%%%%%%%%%%%%%%%%

\subsection{AdS Vacua}

%%%%%%%%%%%%%%%%%%%%%%%%%%%%%%%%%%%%%%%%%%%%%%%%%%%%%%%%%%%%%%%%%%%%%%%%%%%%%%%%%%%%%%%%%

In this section, we will discuss how cosmic strings from section 3
can be embedded into compactifications with stable moduli. For this
we need to stabilize the remaining moduli in~\eqref{3.1}, that is
$S, T^I$ and $Y$. The moduli $T^I$ and $Y$ can be stabilized by the same methods
as in~\cite{BO, Raise}.
However, stabilization of $S$ represents a problem. The imaginary part
of $S$ is the axion. It has been known that cosmic strings of the type constructed
in the previous section are boundaries of axion domain walls~\cite{Wittencosmic}.
If the axion gets a mass the strings become unstable~\cite{Vilenkin}.
This problem can be avoided if the axion is charged under some $U(1)$~\cite{CMP}.
In heterotic M-theory such a situation occurs if there is an anomalous $U(1)$ in
four dimensions. The anomaly of such a $U(1)$ is canceled by the four-dimensional
version of the Green-Schwarz mechanism. The axion, in this case, transforms under $U(1)$ as
\begin{equation}
\delta \sigma =\lambda.
\label{4.1}
\end{equation}
Now the axion becomes a pure gauge degree of freedom and can be
gauged away. As a result, no axion domain walls are formed. The
above arguments will become invalid if the axion receives a
potential. This, in particular, indicates that stability of
cosmic strings requires that no superpotentials for the $S$
modulus should be present. An anomalous $U(1)$ whose presence, as
we just explained, is important for stability of cosmic strings,
gives rise to the moduli dependent FI-terms~\cite{DSRW}. In
heterotic M-theory they are of the form~\cite{Raise}
\begin{equation}
\frac{U_{FI}}{M^{4}_{Pl}}  \sim \frac{g^2}{V^2}.
\label{4.2}
\end{equation}
Here $g$ is the gauge coupling constant which is itself moduli dependent
\begin{equation}
g^2 =\frac{g_0^2}{Re(S+ \dots)},
\label{4.3}
\end{equation}
and the ellipsis stands for the $T^I$- and $Y$-dependent corrections.
The precise form of the corrections depends whether an anomalous $U(1)$ is
in the visible or in the hidden sector. They can be found, for example,
in~\cite{Nonstandard}. To simplify equations below, we will assume that they are
small and can be ignored. Their presence can be shown not to lead to any
conceptually new results below. The order of magnitude of $U_{FI}$
was estimated in~\cite{Raise} and was found to be, generically of order
$\frac{W_f^2}{M^2_{Pl}}$.

As we said, the moduli $T^I$ and $Y$ can be stabilized by methods
presented in~\cite{BO, Raise}. Let us briefly discuss them with a
slight generalization. For simplicity, we will consider the case
$h^{1,1}=1$. Generalizations for $h^{1,1}>1$ is straightforward
but technically more complicated. In this case, there is only one
$T$-modulus whose real part is the length of the interval. If the
five-brane whose modulus has been denoted by $Y$ is located close
to the hidden sector, the leading contribution to the
superpotential $W_0$ is of the form
\begin{equation}
W_0=W_f+\beta e^{-\tau(T-Y)}. \label{n1} \end{equation}
The scale of the coefficient $\beta$ will be assumed to be set by
the eleven-dimensional Planck scale. The equation of motion for
$R$, schematically, is~\cite{BO, Raise}
\begin{equation}
e^{-\tau(R-y)} \sim \frac{|W_f|}{\beta \tau}. \label{n2}
\end{equation}
In order this equation to have a solution, the right hand side
has to be less than one. First, $W_f$ is quantized in units of
$(\frac{\kappa_{11}}{4\pi})^{2/3}$. Therefore, in the limit of
large volume and large interval, the ratio $\frac{|W_f|}{\beta }$
is less than one. Second, the order of magnitude of $W_f$ might be
reduced by Chern-Simons invariants~\cite{Gukovhet}. Third, $\tau$
is much greater than one. This guarantees that, generically, the
right hand side in eq.~\eqref{n2} is less than one. However, to
stabilize $R$ in the phenomenological regime, $R \sim 1$, the
five-brane $Y$ has to be stabilized close to the hidden sector so
that $R-y$ is sufficiently small. Whether or not $Y$ can be
stabilized close enough depends on details of the
compactification. However, there is one obvious possibility of
making $R-y$ small, which has not been considered before. Let us
assume that in the bulk there are $N$ five-branes wrapped on the
same holomorphic curve. Open membrane instantions will generate
contributions to the superpotential inverse proportional to the
exponent of the length between the five-branes. Then the equation
of motion for the $i$-th five-brane will, schematically, be as
follows
\begin{equation}
-\beta_i e^{-\tau (y_i-y_{i-1})}+\beta_{i+1}  e^{-\tau
(y_{i+1}-y_{i})} \sim \frac{|W_f|}{\tau}. \label{n3}
\end{equation}
Writing similar equations for all $N$ five-branes and the
interval, it is possible to show that the five-branes and the
interval will be stabilized in such a way that the distance
between any two adjacent objects will approximately be the same
and, therefore, of order $\frac{R}{N}$. Taking $N$ big enough, it
is always possible to find a solution
\begin{equation}
R \sim 1. \label{n4}
\end{equation}
from eq.~\eqref{n2}.

Let us point out that our analysis in section 3 and in the present
section shows that a five-brain in the bulk might have several
different minima. In particular, there is a minimum close to the
visible sector whose existence is due to the vector bundle moduli.
On the other hand, there is a minimum at a generic point in the
interval due to the non-perturbative exponential interaction with
the adjacent branes (five-branes or the end-of -the-world branes)
on the left and on the right. Putting $N$ five-branes will make
the distance between any two adjacent branes of order
$\frac{R}{N}$.

However, stabilization of the volume becomes a problem.
In~\cite{BO, Gukovhet, Raise}, the volume was stabilized by
balancing the gaugino condensation superpotential against the
fluxes. As we pointed out, any volume dependent superpotential is
also a potential for the axion which will destabilize cosmic
strings. Thus, in this paper, we will assume that no
superpotential for the $S$-modulus is generated. In this
subsection, we will consider the simplest potential for the
volume generated by the Kahler potential and by the FI-terms.
Note that existence of the FI-terms is not an extra assumption. It
already follows from existence of anomalous $U(1)$ which is
necessary to gauge away the axion. The Kahler potential for $S$
is as follows~\cite{Lukas4, Lukas5, Deren}
\begin{equation}
K(S)=-\ln(S+\bar S) +\frac{c}{S+\bar S},
\label{4.4}
\end{equation}
where $c$ depends on $T^I$- and $Y$-moduli which are assumed to
be stable in this paper. To keep the metric in the $S, T^I, Y$
moduli space positive definite, at least in the large volume and
the large interval limit, $c$ has to be sufficiently small so that
the second term is sub-leading. Then by studying equations of
motion for $V$, it is possible to realize that the second term
in~\eqref{4.4} does not play an important role in existence of
extrema in the $V$-direction. Thus, for simplicity, we can assume
that $c$ is sufficiently small and can be ignored to the leading
order. It is also possible to show that the Kahler covariant
derivatives for Kahler and five-brane moduli will be shifted from
zero by terms proportional to $\frac{1}{\tau}$ which is much less
than one. Thus, approximately, dynamics of $V$ is governed by the
potential\footnote{The FI terms will modify equations of motion
for the interval and the five-branes. However, since the order of
magnitude of FI terms is the same as that of the fluxes, a
solution for these moduli will still exist.}
\begin{equation}
U(V) =e^{K(S)} [G^{-1}_{S \bar S} D_S W D_{\bar S} \bar W -3 W \bar W]+U_{FI}
\approx -\frac{A}{V}+\frac{B}{V^3}.
\label{4.5}
\end{equation}
Here $A$ and $B$ are constants (depending on vevs. of other
moduli). The scale of $A$ is set by the fluxes and the scale of
$B$ is set by the FI-terms. Generically, both $A$ and $B$ are of
the same order of magnitude~\cite{Raise}. This potential has a
non-supersymmetric AdS minimum. See fig. 1. Our discussion in
this section shows that it is conceivable to find stable cosmic
strings in heterotic compactifications with a stable
non-supersymmetric AdS vacuum.
\begin{figure}
\epsfxsize=3.5in \epsffile{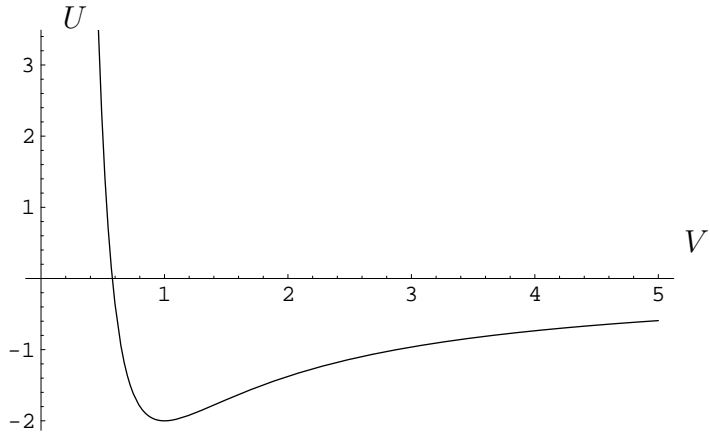}
\begin{picture}(30,30)
\put(-235,155){$U$}
\put(0, 70){$V$}
\end{picture}
\caption{Potential $U(V)$ (appropriately normalized) for $A=3$,
$B=1$. There is an AdS minimum.}
\end{figure}

To conclude this subsection, let us make sure that addition of
new moduli does not destabilize the moduli $\phi$ and $y_0$ from
section 3. It is easy to show that up to terms linear in $y_0$ the
equation of motion for $\phi_i$ is still
\begin{equation}
D_{\phi_i}W=0. \label{4.7}
\end{equation}
Similarly, taking into account the precise dependence of the
FI-terms on $y_0$~\cite{Nonstandard, Raise}, one can show that
eq. $D_{Y_0}W=0$ is modified by terms linear in $y_0$. This means
that the form of eqs.~\eqref{3.10} and~\eqref{3.11} will not
change and a solution $y_0 <<1$ will still exist.

%%%%%%%%%%%%%%%%%%%%%%%%%%%%%%%%%%%%%%%%%%%%%%%%%%%%%%%%%%%%%%%%%%%%%%%%%%%%%%%%%%%%%%%%%%%%%%%%%%%%%%%%%%%%%%%%%%%%%%%%%%%%%%%

\subsection{dS Vacua?}

%%%%%%%%%%%%%%%%%%%%%%%%%%%%%%%%%%%%%%%%%%%%%%%%%%%%%%%%%%%%%%%%%%%%%%%%%%%%%%%%%%%%%%%%%%

Let us now discuss how these results can be extended to
compactifications with dS vacua. The most common strategy to
obtain dS vacua~\cite{KKLT, BKQ, Raise} is to use the F-terms to
obtain a supersymmetric AdS vacuum and then raise with extra terms
(breaking supersymmetry) in the potential energy. However, as we
discussed in the previous subsection, in the presence of cosmic
strings it is problematic to obtain a supersymmetric AdS vacuum
because we cannot use a superpotential for the volume multiplet.
As a result, we had to use FI-terms to stabilize the volume
rather than to raise an AdS vacuum. One way to create dS vacua
can be to stabilize the volume by higher order corrections to the
Kahler potential. This was studied recently in type IIB
compactifications in~\cite{Vijay1, Bobkov, Vijay2}. However, it is
not known how to extend these results to M-theory. One can also
worry that, generically, this approach will stabilize the volume
away from a phenomenologically interesting range. Another approach
could be to search for extra corrections to the potential, in
addition to the F-terms and the FI-terms. Such corrections arise,
for example, in non-supersymmetric compactifications, which, in
general, represent an approach for obtaining dS
vacua~\cite{Raise, Saltman}.

Below, we will show that under some conditions a dS vacuum can be
found in non-supersymmetric $E_8 \times \bar E_8$
compactifications. In such compactifications, there is an extra
volume-dependent contribution to the potential of the
form~\cite{Raise}
\begin{equation}
\Delta U(V) =\frac{a}{V}.
\label{4.8}
\end{equation}
The coefficient $a$ depends on the $h^{1,1}$ moduli and on the
interval. Generically, its order of magnitude is bigger than that
of the fluxes and the FI-terms. However, it depends on the
$h^{1,1}$ moduli. Here, we will assume that the $h^{1,1}$ moduli
are stabilized in such a way that $\Delta U(V)$ is comparable
with $U(V)$. It would be interesting to understand under what
circumstances it is possible to stabilize a two-cycle at a very
small size. We will not discuss it in this paper. Let us consider
the potential $U(V) +\Delta U(V)$ including linear terms in
$\frac{c}{V}$. It looks as follows
\begin{equation}
U(V) +\Delta U(V) =\frac{A^{\prime}}{V}-\frac{C}{V^2}+\frac{B}{V^3}.
\label{4.9}
\end{equation}
Here $A^{\prime}$ is $a-A$. We will assume that it is positive.
$C$ is linear in $c$. It is always positive and comes from the factor
$e^{K(S)}$ in eq.~\eqref{4.5}. It is obvious that under some conditions on the
parameters $A^{\prime}, B$ and $C$, this potential can have a metastable dS vacuum.
See fig. 2.
\begin{figure}
\epsfxsize=3in \epsffile{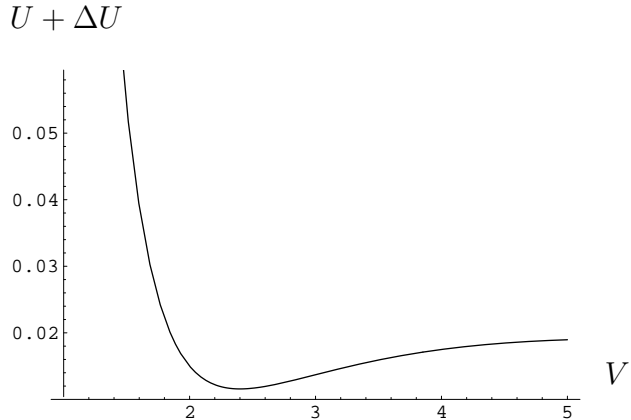}
\begin{picture}(30,30)
\put(-220,150){$U+\Delta U$}
\put(5, 15){$V$}
\end{picture}
\caption{Potential $U(V)+\Delta U(V)$ (appropriately normalized) for
$A=0.25$, $B=1.2$, $C=1$. There is a dS minimum.}
\end{figure}
%

%%%%%%%%%%%%%%%%%%%%%%%%%%%%%%%%%%%%%%%%%%%%%%%%%%%%%%%%%%%%%%%%%%%%%%%%%%%%%%%%%%%%%%%%%%%%%%%%%%%%%%%%%%%%%%%%%%%%%%%%%%

\section{Inflation}

%%%%%%%%%%%%%%%%%%%%%%%%%%%%%%%%%%%%%%%%%%%%%%%%%%%%%%%%%%%%%%%%%%%%%%%%%%%%%%%%%%%%%%%%%%%%%%%%%%%%%%%%%%%%%%%%%%%%%%

In this section, we will briefly discuss whether cosmic strings
studied in this paper can be formed after inflation. For
concreteness, we will consider the inflationary model
of~\cite{Myinflation}. One should also be able to extend this
discussion to another heterotic M-theory inflation studied
in~\cite{Theirinflation}.

To the system of moduli in eq.~\eqref{3.1}, we add another
five-brane multiplet corresponding to a five-brane wrapped on a
non-isolated genus zero or higher genus holomorphic curve. Moduli
of such a five-brane will not have any non-perturbative
superpotential. Instead, the translational modulus has a
potential with inflationary
properties\footnote{In~\cite{Myinflation}, the extra five-brane
multiplet had a superpotential due to a gaugino condensate in the
hidden sector. In this paper, we are assuming that this
superpotential is zero because it would also be a potential for
the axion. However, performing the same analysis as
in~\cite{Myinflation}, it is not hard to show that in the absence
of the gaugino condensation superpotential, the potential for the
inflaton will be exactly of the same form as
in~\cite{Myinflation}.}. The process of inflation is represented
by such a five-brane approaching the visible brane. It was argued
in~\cite{Myinflation} that when the five-brane comes close to the
visible sector, the transition vector bundle moduli~\cite{BDO1,
Rene} become tachyonic and terminate inflation. Eventually, the
five-brane disappears through the small instanton transition
which modifies the vector bundle on the visible brane, in
particular, new vector bundle moduli are created which can be
stabilized by non-perturbative effects. Thus, formation of
macroscopic open membranes with a small length should be related
to this change of the vector bundle.

Since after the small instanton transition, the vector bundle,
the number of the vector bundle moduli and the Pfaffian change,
the position of the five-brane discussed in section 3 will also
change. This could explain why it is stabilized close to the
visible sector only after inflation. For example, let the bundle
before inflation not have any moduli. Then from discussion in
section 3 it follows that no minimum for the five-brane close to
the visible sector will exist. Therefore, in this case, the
five-brane will be stabilized at a generic point along the
interval. After inflation, vector bundle moduli will be created
and, according to our discussion in section 3, they might force
the five-brane to move closer to the visible sector so that
eq.~\eqref{1.2} is satisfied. The presence of vector bundle
moduli might also stabilize a five-brane close to the visible
brane after inflation in the model proposed
in~\cite{Theirinflation}. Thus, it might be possible to create
cosmic strings in the inflationary scenario
in~\cite{Theirinflation}. It would be interesting to understand
it in detail.

%%%%%%%%%%%%%%%%%%%%%%%%%%%%%%%%%%%%%%%%%%%%%%%%%%%%%%%%%%%%%%%%%%%%%%%%%%%%%%%%%%%%%%%%%%%%%%%%%%%%%%%%%%%%%%%%%%%%%%%%%%%%

\section{Acknowledgments}

%%%%%%%%%%%%%%%%%%%%%%%%%%%%%%%%%%%%%%%%%%%%%%%%%%%%%%%%%%%%%%%%%%%%%%%%%%%%%%%%%%%%%%%%%%%%%%%%%%%%%%%%%%%%%%
The author is very grateful to Juan Maldacena for lots of helpful
discussions, explanations and comments on the preliminary version
of the paper. The author would also like to thank Willy Fischler
for discussions. The work is supported by NSF grant PHY-0070928.

%%%%%%%%%%%%%%%%%%%%%%%%%%%%%%%%%%%%%%%%%%%%%%%%%%%%%%%%%%%%%%%%%%%%%%%%%%%%%%%%%%%%%%%%%%%%%%%%%%%%%%%%%%%%%%%%%%%%%

%%%%%%%%%%%%%%%%%%%%%%%%%%%%%%%%%%%%%%%%%%%%%%%%%%%%%%%%%%%%%%%%%%%%%%%%%%%%%%%%%%%%%%%%%%%%%%%%%%%%%%%%%%%%%%

\end{document}